\documentclass[conference]{IEEEtran}

\usepackage{enumerate}
\usepackage{amsmath}
\usepackage{amssymb}
\usepackage{amsthm}
\usepackage{cite}

\newtheorem{theorem}{Theorem}[section]
\newtheorem{proposition}[theorem]{Proposition}

\newcommand{\supp}{\textnormal{supp}}

\newcommand{\abs}[1]{\lvert#1\rvert} 
\newcommand{\card}[1]{\abs{#1}} 
\newcommand{\comp}[1]{{#1}^{\textnormal{c}}} 
\newcommand{\set}[1]{\mathcal{#1}}

\newcommand{\bigE}[1]{\operatorname{E}\bigl[#1\bigr]} 

\newcommand{\rhotrans}{\tilde{\rho}}
\newcommand{\naturals}{\mathbb{N}}

\IEEEoverridecommandlockouts

\begin{document}

\sloppy

\title{Codes for Tasks and R\'enyi Entropy Rate}

\author{
  \IEEEauthorblockN{Christoph Bunte and Amos Lapidoth}
  \IEEEauthorblockA{ETH Zurich\\
    Email: \{bunte,lapidoth\}@isi.ee.ethz.ch} 
}



\maketitle

\begin{abstract}
A task is randomly drawn from a finite set of tasks and is described using a fixed
number of bits. All the tasks that share its description must be performed. Upper and
lower bounds on the minimum $\rho$-th moment of the number of performed tasks 
are derived. The key is an analog of the Kraft Inequality for partitions of
finite sets. 
When a sequence of tasks is produced by a source of a given
R\'enyi entropy rate of order $1/(1+\rho)$ and~$n$ tasks are jointly described 
using~$nR$ bits, it is shown that for~$R$ larger than the R\'enyi entropy rate, 
the $\rho$-th moment of the ratio of performed tasks to $n$ 
can be driven to one as $n$ tends to infinity, and that for $R$ less than the
R\'enyi entropy rate it tends to infinity. 
This generalizes a recent result for IID sources by the same authors. 
A mismatched version of the direct part is also considered, where the code is
designed according to the wrong law. The penalty incurred by the mismatch can be
expressed in terms of a divergence measure that was shown by Sundaresan 
to play a similar role in the Massey-Arikan guessing problem.
\end{abstract}

\section{Introduction}
\label{sec:introduction}
You are asked to complete a task $X$ drawn according to a PMF $P$ 
from a finite set of tasks $\set{X}$. You do not get to see $X$ but only its
description $f(X)$, where 
\begin{equation}
\label{eq:encoder}
f\colon\set{X}\to\{1,\ldots,M\}.
\end{equation}
In other words, $X$ is described to you using $\log M$ bits. 
You know the mapping $f$ and you
promise to complete $X$ based on $f(X)$, which leaves you no choice but to
complete every task in the set
\begin{equation}
f^{-1}(f(X)) = \{x\in\set{X}: f(x)=f(X)\}.
\end{equation}
In the interesting case where $M < \card{\set{X}}$, you will sometimes have to
perform multiple tasks, of which all but one are superfluous. (We use
$\card{\cdot}$ to denote the cardinality of sets.)

Given $M$, the goal is to design $f$ so as to
minimize the $\rho$-th moment of the number of tasks you perform
\begin{equation}
\label{eq:rhoth_moment}
\bigE{\card{f^{-1}(f(X))}^\rho} = \sum_{x\in\set{X}} P(x)
\card{f^{-1}(f(x))}^\rho,
\end{equation}
where $\rho$ is some given positive number. This minimum is at least one 
because~$X$ is in~$f^{-1}(f(X))$; it decreases as $M$ increases; and it is equal to one when $M \geq
\card{\set{X}}$. 

Our first result is a pair of upper and lower bounds on this 
minimum as a function of $M$. The bounds 
are expressed in terms of the \emph{R\'enyi entropy of $X$ of
order $1/(1+\rho)$}:
\begin{equation}
\label{eq:renyi}
H_{\frac{1}{1+\rho}}(X) = \frac{1+\rho}{\rho} \log
\sum_{x\in\set{X}}P(x)^{\frac{1}{1+\rho}}.
\end{equation}
Throughout $\log(\cdot)$ stands for $\log_2(\cdot)$, the logarithm to base $2$. 
For typographic reasons we henceforth use the notation
\begin{equation}
\label{eq:rhotrans}
\rhotrans=\frac{1}{1+\rho},\quad \rho>0.
\end{equation}
\begin{theorem}
\label{thm:oneshot}
Let $\rho>0$. 
\begin{enumerate}
\item For all positive integers $M$ and every $f\colon\set{X} \to
\{1,\ldots,M\}$, 
\begin{equation}
\bigE{\card{f^{-1}(f(X))}^\rho}
\geq 2^{\rho(H_{\rhotrans}(X)-\log M)}.
\end{equation}
\item For every integer $M > \log\card{\set{X}}+2$ there exists $f\colon
\set{X}\to\{1,\ldots,M\}$ such that
\begin{equation}
\bigE{\card{f^{-1}(f(X))}^\rho}\\
<1+2^{\rho(H_{\rhotrans}(X)-\log \widetilde{M})},
\end{equation}
where $\widetilde{M}=(M-\log\card{\set{X}}-2)/4$.
\end{enumerate}
\end{theorem}
A proof is provided in Section~\ref{sec:proof}.
The lower bound is essentially~\cite[Lemma III.1]{bunte2013source}.

Theorem~\ref{thm:oneshot} is particularly useful when applied to the case where a sequence of tasks is produced by a
source~$\{X_i\}_{i=1}^\infty$ with alphabet~$\set{X}$ and the first $n$ tasks
$X^n=(X_1,\ldots,X_n)$ are jointly described using~$nR$ bits:
\begin{equation}
f \colon\set{X}^n \to \{1,\ldots, 2^{nR}\}.
\end{equation}
We assume that the order in which the tasks are performed matters 
and that every $n$-tuple of tasks in
the set $f^{-1}(f(X^n))$ must be performed. The total number of performed tasks
is therefore $n \card{f^{-1}(f(X^n))}$, and the ratio of the number of performed
tasks to the number of assigned tasks is $\card{f^{-1}(f(X^n))}$. 
\begin{theorem}
\label{thm:main}
Let $\{X_i\}_{i=1}^\infty$ be any source with finite alphabet~$\set{X}$.
\begin{enumerate}
\item If $R>\limsup_{n\to\infty} H_{\rhotrans}(X^n)/n$, 
then there exist encoders $f_n\colon \set{X}^n \to \{1,\ldots, 
2^{nR}\}$ such that\footnote{Throughout $2^{nR}$ stands for $\lfloor
2^{nR}\rfloor$.}
\begin{equation}
\lim_{n\to\infty}\bigE{\card{f^{-1}_n(f_n(X^n))}^\rho} =1.
\end{equation}
\item If $R<\liminf_{n\to\infty} H_{\rhotrans}(X^n)/n$, 
then for any choice of encoders $f_n \colon \set{X}^n \to \{1,\ldots,2^{nR}\}$, 
\begin{equation}
\lim_{n\to\infty}\bigE{\card{f^{-1}_n(f_n(X^n))}^\rho} =\infty.
\end{equation}
\end{enumerate}
\end{theorem}
\begin{IEEEproof}
On account of Theorem~\ref{thm:oneshot}, for all $n$ large enough so that
$2^{nR} > n \log
\card{\set{X}}+2$, 
\begin{multline}
2^{n\rho\bigl(\frac{H_{\rhotrans}(X^n)}{n}- R\bigr)} \leq \min_{f_n\colon \set{X}^n \to
\{1,\ldots,2^{nR}\}} \bigE{\card{f_n^{-1}(f_n(X^n))}^\rho}\\ < 1+
2^{n\rho\bigl(\frac{H_{\rhotrans}(X^n)}{n} - R+\delta_n\bigr)},
\end{multline}
where $\delta_n \to 0$ as $n\to\infty$. 
\end{IEEEproof}
When it exists, the limit 
\begin{equation}
\lim_{n\to\infty} \frac{H_{\alpha}(X^n)}{n}
\end{equation}
is called the \emph{R\'enyi entropy rate of order $\alpha$}.
It exists for a large class of sources, including time-invariant Markov
sources~\cite{rached2001renyi,pfister2004renyi,malone2004guesswork}. 
Theorem~\ref{thm:main} generalizes \cite[Theorem IV.1]{bunte2013source} from IID
sources to sources with memory and furnishes an operational characterization of the R\'enyi entropy
rate for all orders in~$(0,1)$. 
Note that for IID sources the R\'enyi entropy rate reduces to the R\'enyi entropy because
in this case~$H_{\rhotrans}(X^n) = n H_{\rhotrans}(X_1)$.

The proof of the lower bound in Theorem~\ref{thm:oneshot} hinges on the
following simple observation. 
\begin{proposition}
\label{prop:count_lists}
If $\set{L}_1,\ldots,\set{L}_M$ is a partition of a finite 
set~$\set{X}$ into $M$ nonempty subsets (i.e., $\bigcup_{m=1}^M \set{L}_m =
\set{X}$ and $\set{L}_m\cap\set{L}_{m'} = \emptyset$ if, and only if,  $m'\neq
m$), and $L(x)$ is the cardinality of the
subset containing $x$, then
\begin{equation}
\label{eq:count_lists}
\sum_{x\in\set{X}} \frac{1}{L(x)} = M. 
\end{equation}
\end{proposition}
\begin{IEEEproof}
\begin{align}
\sum_{x\in\set{X}} \frac{1}{L(x)}&= \sum_{m=1}^M \sum_{x\in\set{L}_m}
\frac{1}{L(x)}\\
&=\sum_{m=1}^M \sum_{x\in\set{L}_m} \frac{1}{\card{\set{L}_m}}\\
&= M.
\end{align}
\end{IEEEproof}
Note that the reverse of Proposition~\ref{prop:count_lists} 
is not true in the sense that if $\lambda \colon \set{X}
\to \naturals\triangleq\{1,2,\ldots\}$ satisfies
\begin{equation}
\sum_{x\in\set{X}} \frac{1}{\lambda(x)} = \mu,
\end{equation}
then there need not exist a partition of $\set{X}$ into $\lceil\mu\rceil$ subsets
such that the cardinality of the subset containing $x$ is at most~$\lambda(x)$. 
A counterexample is $\set{X}=\{a,b,c,d\}$ with $\lambda(a)=1$,
$\lambda(b)=2$, and $\lambda(c)=\lambda(d)=4$. In this example, $\mu=2$, but we need~3
subsets to satisfy the cardinality constraints. 

However, as our next result shows, allowing a slightly larger number of subsets suffices:
\begin{proposition}
\label{prop:sufficiency}
If $\set{X}$ is a finite set, $\lambda\colon \set{X} \to \naturals\cup\{+\infty\}$ and
\begin{equation}
\sum_{x\in\set{X}} \frac{1}{\lambda(x)} = \mu
\end{equation}
(with the convention $1/\infty=0$), then there exists a partition of $\set{X}$ into at most
\begin{equation}
\label{eq:alpha}
\min_{\alpha >1} \lfloor\alpha \mu + \log_\alpha \card{\set{X}}+2\rfloor
\end{equation}
subsets such that
\begin{equation}
\label{eq:card_bound}
L(x) \leq \min\{\lambda(x),\card{\set{X}}\},\quad \text{for all $x\in\set{X}$,}
\end{equation}
where $L(x)$ is the cardinality of the subset containing $x$. 
\end{proposition}
Proposition~\ref{prop:sufficiency} is the key to the upper bound
in Theorem~\ref{thm:oneshot}. Combined with Proposition~\ref{prop:count_lists}
it can be considered an analog of the Kraft Inequality \cite[Theorem 5.5.1]{cover2006elements} 
for partitions of finite sets. 
A proof is given in Section~\ref{sec:proposition}.

The construction of the encoder in the derivation of the upper bound in
Theorem~\ref{thm:oneshot} requires knowledge of the distribution $P$ of $X$ (see
Section~\ref{sec:upper_bound}).
In Section~\ref{sec:divergence} we consider a mismatched version of this direct
part where the construction is carried out based on the law $Q$ instead of $P$. 
We show that the penalty incurred by the mismatch between $P$ and $Q$ can be expressed in
terms of the divergence measures
\begin{equation}
\label{eq:divergence}
\Delta_\alpha(P||Q) \triangleq \log \frac{\sum_{x\in\set{X}}
Q(x)^\alpha}{\bigl(\sum_{x\in\set{X}} P(x)^\alpha\bigr)^{\frac{1}{1-\alpha}}}\biggl(\sum_{x\in\set{X}}
\frac{P(x)}{Q(x)^{1-\alpha}}\biggr)^{\frac{\alpha}{1-\alpha}},
\end{equation}
where $\alpha$ can be any positive number not equal to one. (We use the convention $0/0=0$ and
$a/0=+\infty$ if $a>0$.) 
This family of divergence measures was proposed by
Sundaresan~\cite{sundaresan2007guessing}, who showed that it plays a similar
role in the Massey-Arikan guessing problem~\cite{massey1994guessing, arikan1996inequality}.

\section{Proof of Theorem~\ref{thm:oneshot}}
\label{sec:proof}
\subsection{The Lower Bound (Converse)}
The proof of the lower bound is inspired by the proof of~\cite[Theorem 1]{arikan1996inequality}.
Fix an encoder $f\colon\set{X}\to\{1,\ldots,M\}$, 
and note that it gives rise to a partition of~$\set{X}$ into the $M$ subsets 
\begin{equation}
\label{eq:induced_partition2}
\{x\in\set{X}:f(x)=m\},\quad  m\in \{1,\ldots, M\}.
\end{equation}
Let $N$ denote the number of nonempty subsets
in this partition.
Also note that for this partition the cardinality of the subset containing $x$
is 
\begin{equation}
\label{eq:induced_partition}
L(x) = \card{f^{-1}(f(x))},\quad \text{for all $x\in\set{X}$.}
\end{equation}
Recall H\"older's Inequality: If $a,b\colon \set{X} \to [0,\infty)$, $p,q>1$ and $1/p+1/q=1$, then 
\begin{equation}
\label{eq:hoelder}
\sum_{x\in\set{X}} a(x) b(x) \leq \biggl(\sum_{x\in\set{X}} a(x)^p\biggr)^{1/p}
\biggl(\sum_{x\in\set{X}} b(x)^q\biggr)^{1/q}.
\end{equation}
Rearranging~\eqref{eq:hoelder} gives
\begin{equation}
\label{eq:hoelder_alt}
\sum_{x\in\set{X}} a(x)^p \geq \biggl(\sum_{x\in\set{X}} b(x)^q\biggr)^{-p/q}
\biggl(\sum_{x\in\set{X}} a(x)b(x)\biggr)^{p}.
\end{equation}
Substituting $p=1+\rho$, $q=(1+\rho)/\rho$, $a(x) = P(x)^{\frac{1}{1+\rho}}
\card{f^{-1}(f(x))}^{\frac{\rho}{1+\rho}}$ and $b(x) =
\card{f^{-1}(f(x))}^{-\frac{\rho}{1+\rho}}$ in~\eqref{eq:hoelder_alt}, we obtain
\begin{align}
&\sum_{x\in\set{X}} P(x) \card{f^{-1}(f(x))}^{\rho}\\ 
&\quad\geq \biggl(\sum_{x\in\set{X}} \frac{1}{\card{f^{-1}(f(x))}}\biggr)^{-\rho}
\biggl(\sum_{x\in\set{X}} P(x)^{\frac{1}{1+\rho}}
\biggr)^{1+\rho}\\
&\quad=  2^{\rho(H_{\rhotrans}(X)-\log N)}\label{eq:440_43}\\
&\quad\geq  2^{\rho(H_{\rhotrans}(X)-\log M)},\label{eq:441_532}
\end{align}
where~\eqref{eq:440_43} follows from~\eqref{eq:renyi}, \eqref{eq:induced_partition}, and
Proposition~\ref{prop:count_lists}; and where~\eqref{eq:441_532} follows
because $N \leq M$.\hfill\IEEEQED
\subsection{The Upper Bound (Direct Part)}
\label{sec:upper_bound}
Since H\"older's Inequality~\eqref{eq:hoelder} holds with equality if, and only
if, (iff) $a(x)^p$ is
proportional to $b(x)^q$, it follows that the
lower bound in Theorem~\ref{thm:oneshot} holds with equality 
iff $\card{f^{-1}(f(x))}$ is proportional to $P(x)^{-1/(1+\rho)}$. We derive the
upper bound in Theorem~\ref{thm:oneshot} by constructing a partition that approximately satisfies this
relationship.
To this end, we use Proposition~\ref{prop:sufficiency} with $\alpha=2$ in~\eqref{eq:alpha}
and
\begin{equation}
\label{eq:lambda}
\lambda(x) = \begin{cases} \bigl\lceil \beta\,
P(x)^{-\frac{1}{1+\rho}}\bigr\rceil& \text{if $P(x)>0$,}\\+\infty& \text{if
$P(x)=0$,}\end{cases}
\end{equation}
where we choose $\beta$ just large enough to guarantee the existence of a
partition of $\set{X}$ into at most $M$ subsets  
satisfying~\eqref{eq:card_bound}. This is accomplished by the choice
\begin{equation}
\label{eq:beta}
\beta= \frac{2\sum_{x\in\set{X}}
P(x)^{\frac{1}{1+\rho}}}{M-\log\card{\set{X}}-2}.
\end{equation}
(This is where we need $M> \log\card{\set{X}}+2$.) 
Indeed, 
\begin{align}
\mu&=\sum_{x\in\set{X}} \frac{1}{\lambda(x)} \\
&\leq \sum_{x\in\set{X}} \frac{P(x)^{\frac{1}{1+\rho}}}{\beta} \\
&=\frac{M-\log\card{\set{X}}-2}{2},
\end{align}
and hence
\begin{equation}
2\mu + \log \card{\set{X}} +2 \leq M.
\end{equation}
Let then the partition
$\set{L}_1,\ldots,\set{L}_N$ with $N \leq M$ be as promised by
Proposition~\ref{prop:sufficiency}, and  
construct $f\colon\set{X}\to\{1,\ldots,M\}$ by setting $f(x) = m$ if $x\in
\set{L}_m$. For this encoder, 
\begin{align}
\sum_{x\in\set{X}} P(x)\card{f^{-1}(f(x))}^\rho &= \sum_{x:P(x)>0} P(x)
L(x)^\rho\label{eq:370}\\
&\leq \sum_{x:P(x)>0} P(x) \lambda(x)^\rho\label{eq:371}\\
&<1+ 2^{\rho(H_{\rhotrans}(X)-\log \widetilde{M})},\label{eq:372}
\end{align}
where the strict inequality follows from~\eqref{eq:lambda} and the inequality
\begin{equation}
\lceil\xi\rceil^\rho<1+2^\rho \xi^\rho,\quad \text{for all $\xi\geq0$,}
\end{equation}
which is easily checked by considering separately the cases~$0\leq\xi\leq 1$
and~$\xi>1$. \hfill\IEEEQED

\section{Proof of Proposition~\ref{prop:sufficiency}}
\label{sec:proposition}
We describe a procedure for constructing a partition of~$\set{X}$ with the desired properties. 
Since the labels do not matter, we may assume for convenience of notation that
$\set{X}=\{1,\ldots,\card{\set{X}}\}$ and 
\begin{equation}
\label{eq:increasing}
\lambda(1) \leq \lambda(2) \leq \cdots \leq \lambda(\card{\set{X}}).
\end{equation}
The first subset in the partition we construct is
\begin{equation}
\label{eq:L0}
\set{L}_0=\{x\in\set{X}:\lambda(x) \geq \card{\set{X}}\}. 
\end{equation}
If $\set{X}=\set{L}_0$, then the construction is complete and~\eqref{eq:alpha} and~\eqref{eq:card_bound} are 
clearly satisfied. Otherwise we follow the steps below to construct additional
subsets
$\set{L}_1,\ldots,\set{L}_M$.
\begin{quote}
\emph{Step $1$}: If 
\begin{equation}
\card{\set{X} \setminus \set{L}_0} \leq
\lambda(1), 
\end{equation}
then we complete the construction by setting 
$\set{L}_{1} = \set{X}\setminus\set{L}_0$ and $M=1$. Otherwise 
we set
\begin{equation}
\set{L}_1 = \bigl\{1,\ldots,\lambda(1)\bigr\}
\end{equation}
and go to Step $2$.\\
\emph{Step $m \geq 2$}: If 
\begin{equation}
\biggl|\set{X} \setminus \bigcup_{i=0}^{m-1} \set{L}_i\biggr| \leq
\lambda(\card{\set{L}_1}+\ldots+\card{\set{L}_{m-1}}+1), 
\end{equation}
then we complete the construction by setting $\set{L}_{m} =
\set{X}\setminus\bigcup_{i=0}^{m-1} \set{L}_i$ and $M=m$. Otherwise 
we let $\set{L}_{m}$ contain the
$\lambda(\card{\set{L}_1}+\ldots+\card{\set{L}_{m-1}}+1)$ smallest elements of
$\set{X}\setminus\bigcup_{i=0}^{m-1} \set{L}_i$, i.e., we set
\begin{multline}
\set{L}_m = \bigl\{\card{\set{L}_1}+\ldots+\card{\set{L}_{m-1}}+1,\ldots,\\
\card{\set{L}_1}+\ldots+\card{\set{L}_{m-1}}+\lambda(\card{\set{L}_1}+\ldots+\card{\set{L}_{m-1}}+1)\bigr\}
\end{multline}
and go to Step $m+1$.  
\end{quote}
We next verify that~\eqref{eq:card_bound} is satisfied and that
the total number of subsets $M+1$ does not exceed~\eqref{eq:alpha}. 
Clearly, $L(x) \leq \card{\set{X}}$ for every $x\in \set{X}$, so to
prove~\eqref{eq:card_bound} we check
that $L(x) \leq \lambda(x)$ for every $x\in\set{X}$. 
It is clear that $L(x) \leq
\lambda(x)$ for all $x\in \set{L}_0$. Let $k(x)$ denote the smallest element in
the subset containing $x$. Then $L(x) \leq \lambda(k(x))$ for all $x\in
\bigcup_{m=1}^M \set{L}_m$ by construction, and
since $k(x) \leq x$, we have $\lambda(k(x)) \leq \lambda(x)$ by the
assumption~\eqref{eq:increasing}, and hence $L(x) \leq \lambda(x)$ for all $x\in
\set{X}$. 

It remains to
check that $M+1$ does not exceed~\eqref{eq:alpha}. This is clearly true when
$M=1$, so we assume that $M \geq 2$. 
Since $L(x) = \lambda(k(x))$ for all $x\in \bigcup_{m=1}^{M-1} \set{L}_m$, we
have on account of Proposition~\ref{prop:count_lists} 
\begin{align}
M &=  \sum_{x\in\bigcup_{m=1}^M \set{L}_m} \frac{1}{L(x)}\\
&= 1+\sum_{x\in \bigcup_{m=1}^{M-1}\set{L}_m} \frac{1}{L(x)}\\
&= 1+ \sum_{x\in\bigcup_{m=1}^{M-1}\set{L}_m}
\frac{1}{\lambda(k(x))}.\label{eq:519}
\end{align}
Fix an arbitrary $\alpha>1$ and let $\set{M}$ be the set of indices $m \in \{1,\ldots,M-1\}$ such that there is an
$x\in \set{L}_m$ with $\lambda(x) > \alpha \lambda(k(x))$. We next argue that
$\card{\set{M}} < \log_\alpha \card{\set{X}}$. To this end, 
enumerate the indices in $\set{M}$ as $m_1<m_2 < \cdots < m_{\card{\set{M}}}$.
For each $i\in \{1,\ldots,\card{\set{M}}\}$ select $x_{i}\in \set{L}_{m_i}$ such that
$\lambda(x_{i}) > \alpha \lambda(k(x_i))$. Then 
\begin{align}
\lambda(x_{1}) &> \alpha\lambda(k(x_1))\\
&\geq \alpha.
\end{align}
Note that if $m<m'$ and $x\in\set{L}_m$ and $x' \in \set{L}_{m'}$, then $x<x'$.
Thus, $x_1<k(x_2)$ because $x_1\in\set{L}_{m_1}$ and $k(x_2) \in \set{L}_{m_2}$,
and $m_1<m_2$. Consequently, 
\begin{align}
\label{eq:632_24}
\lambda(x_{2}) &> \alpha \lambda(k(x_{2}))\\
&\geq \alpha \lambda(x_{1})\\
&> \alpha^2. 
\end{align}
Iterating this argument shows that
\begin{align}
\lambda(x_{\card{\set{M}}})> \alpha^{\card{\set{M}}}.
\end{align}
And since $\lambda(x) \leq \card{\set{X}}$ for $x\in \bigcup_{m=1}^M \set{L}_m$ by~\eqref{eq:L0},
it follows that $\card{\set{M}} < \log_\alpha \card{\set{X}}$.
Continuing from~\eqref{eq:519} with $\comp{\set{M}} \triangleq \{1,\ldots, M-1\} \setminus
\set{M}$, 
\begin{align}
M&= 1+ \card{\set{M}} + \sum_{x\in \bigcup_{m\in \comp{\set{M}}}\set{L}_m} \frac{1}{\lambda(k(x))}\\
&< 1 + \log_\alpha \card{\set{X}} + \alpha \sum_{x\in \bigcup_{m\in \comp{\set{M}}}\set{L}_m}
\frac{1}{\lambda(x)} \\
&\leq 1 + \log_\alpha \card{\set{X}} + \alpha \mu,\label{eq:723_final}
\end{align}
where the first inequality follows because $\lambda(x) \leq \alpha
\lambda(k(x))$ for $x\in \bigcup_{m\in \comp{\set{M}}} \set{L}_m$, and where the
second inequality follows from the hypothesis of the proposition. Since $M+1$ is an
integer and $\alpha>1$ is arbitrary, it follows from~\eqref{eq:723_final} that $M+1$ is upper-bounded
by~\eqref{eq:alpha}.\hfill\IEEEQED

\section{Mismatch}
\label{sec:divergence}
The key to the upper bound in Theorem~\ref{thm:oneshot} was to use
Proposition~\ref{prop:sufficiency} with $\lambda$ as in~\eqref{eq:lambda}
and~\eqref{eq:beta} to obtain a partition of $\set{X}$ for which the cardinality of the
subset containing $x$ is approximately proportional to
$P(x)^{-1/(1+\rho)}$. Evidently, this construction requires knowledge of the
distribution~$P$ of~$X$. 
In this section, we derive the penalty when~$P$ is replaced with~$Q$
in~\eqref{eq:lambda} and~\eqref{eq:beta}. 
Since it is then still true that
\begin{equation}
\mu \leq \frac{M-\log \card{\set{X}}-2}{2}, 
\end{equation}
Proposition~\ref{prop:sufficiency} guarantees the existence of a partition 
of~$\set{X}$ into at most~$M$ subsets satisfying~\eqref{eq:card_bound}. Constructing 
$f$ from this partition as in Section~\ref{sec:upper_bound} 
and proceeding similarly as
in~\eqref{eq:370} to~\eqref{eq:372}, we obtain
\begin{equation}
\label{eq:mismatch_bound}
\sum_{x\in\set{X}} P(x)\card{f^{-1}(f(x))}^\rho<1+2^{\rho(H_{\rhotrans}(X) +
\Delta_{\rhotrans}(P||Q) - \log \widetilde{M})},
\end{equation}
where $\Delta_{\rhotrans}(P||Q)$ is as in~\eqref{eq:divergence} and $\widetilde{M}$
is as in Theorem~\ref{thm:oneshot}. (Note that
$\Delta_{\rhotrans}(P||Q)<\infty$ only if the support of $P$ is contained in the
support of $Q$.) The penalty in the exponent when compared to the upper bound in
Theorem~\ref{thm:oneshot} is thus given by $\Delta_{\rhotrans}(P||Q)$. 
To reinforce this, further note that 
\begin{equation}
\label{eq:product_div}
\Delta_\alpha(P^n||Q^n) = n \Delta_\alpha(P||Q),
\end{equation}
where $P^n$ and $Q^n$ are the $n$-fold products of $P$ and $Q$. 
Consequently, if the source $\{X_i\}_{i=1}^\infty$ is IID $P$ and we construct
$f_n\colon\set{X}^n\to\{1,\ldots,2^{nR}\}$ similarly as above based on $Q^n$ instead of $P^n$, 
we obtain the bound
\begin{equation}
\label{eq:mismatch_upper_bound}
\bigE{\card{f_n^{-1}(f_n(X^n))}^\rho}<1+2^{n\rho(H_{\rhotrans}(X_1) +
\Delta_{\rhotrans}(P||Q) - R + \delta_n)},
\end{equation}
where $\delta_n\to 0$ as $n\to\infty$. 
The RHS of~\eqref{eq:mismatch_upper_bound} tends to one provided that
$R>H_{\rhotrans}(X_1)+\Delta_{\rhotrans}(P||Q)$. 
Thus, in the IID case $\Delta_{\rhotrans}(P||Q)$
is the rate penalty incurred by the mismatch between $P$ and $Q$. 
 
We conclude this section with some properties of~$\Delta_{\alpha}(P||Q)$. 
Properties 1--3 (see below) were given in~\cite{sundaresan2007guessing}; we
repeat them here for completeness.
Note that R\'enyi's divergence (see, e.g.,~\cite{csiszar1995generalized})
\begin{equation}
D_\alpha(P||Q) = \frac{1}{\alpha-1} \log \sum_{x\in\set{X}} P(x)^\alpha
Q(x)^{1-\alpha},
\end{equation}
satisfies Properties 1 and 3 but none of the others in general.
\begin{proposition}
\label{prop:properties}
Let $\supp(P)$ and $\supp(Q)$ denote the support sets of $P$ and $Q$.
The functional $\Delta_\alpha(P||Q)$ has the following properties.
\begin{enumerate}
\item $\Delta_\alpha(P||Q) \geq 0$ with equality iff $P=Q$. 
\item $\Delta_\alpha(P||Q)=\infty$ iff ($0<\alpha<1$ and $\supp(P)\not\subseteq
\supp(Q)$) or ($\alpha>1$ and $\supp(P)\cap \supp(Q) = \emptyset$.) 
\item $\lim_{\alpha\to 1} \Delta_\alpha(P||Q) = D(P||Q)$. 
\item $\lim_{\alpha\to 0}\Delta_\alpha(P||Q) = \log
\frac{\card{\supp(Q)}}{\card{\supp(P)}}$ if $\supp(P) \subseteq \supp(Q)$. 
\item $\lim_{\alpha\to \infty} \Delta_\alpha(P||Q) = \log
\frac{\max_{x\in\set{X}}
P(x)}{\frac{1}{\card{\set{Q}}}\sum_{x'\in\set{Q}}P(x')}$, where
\begin{equation*}
\set{Q}=\bigl\{x\in \set{X}: Q(x) = \max_{x'\in\set{X}} Q(x')\bigr\}. 
\end{equation*}
\end{enumerate}
\end{proposition}
\begin{IEEEproof}
Property 2 follows by inspection of~\eqref{eq:divergence}. Properties 3--5 follow by
simple calculus. As to Property 1, 
consider first the case where $0<\alpha<1$. In view of Property 2, we may assume
that $\supp(P) \subseteq \supp(Q)$. H\"older's Inequality~\eqref{eq:hoelder} with
$p=1/\alpha$ and $q=1/(1-\alpha)$ gives
\begin{align}
\sum_{x\in\set{X}} P(x)^\alpha&=\sum_{x\in\supp(P)}
\frac{P(x)^\alpha}{Q(x)^{\alpha(1-\alpha)}} Q(x)^{\alpha(1-\alpha)}\\
&\leq \biggl(\sum_{x\in\supp(P)}
 \frac{P(x)}{Q(x)^{1-\alpha}}\biggr)^\alpha \biggl(\sum_{x\in\supp(P)} Q(x)^\alpha
\biggr)^{1-\alpha}\notag\\
&\leq \biggl(\sum_{x\in\set{X}}
 \frac{P(x)}{Q(x)^{1-\alpha}}\biggr)^\alpha \biggl(\sum_{x\in\set{X}} Q(x)^\alpha
\biggr)^{1-\alpha}.
\end{align}
The conditions for equality in H\"older's Inequality imply that 
equality holds iff $P=Q$. Consider next the 
case where~$\alpha>1$. By H\"older's Inequality with $p=\alpha$ and
$q=\alpha/(\alpha-1)$,
\begin{align}
\sum_{x\in\set{X}}\frac{P(x)}{Q(x)^{1-\alpha}}&=\sum_{x\in\set{X}}P(x)
Q(x)^{\alpha-1}\\
&\leq \biggl(\sum_{x\in\set{X}} P(x)^\alpha\biggr)^{\frac{1}{\alpha}}
\biggl(\sum_{x\in\set{X}} Q(x)^{\alpha}\biggr)^{\frac{\alpha-1}{\alpha}},
\end{align}
with equality iff $P=Q$. 
\end{IEEEproof}




\bibliographystyle{IEEEtran}


\end{document}